\documentclass[12pt]{elsart}
\usepackage{feynmp,epsfig,graphics}

\def\tr{{\rm tr} \,}

\def\pslash{\FMslash p}

\def\qslash{\FMslash q}
\def\barqslash{\FMslash{\bar{q}}}

\def\wslash{\FMslash w}

\begin{document}
GSI-Preprint-2001-18, nucl-th/0105067
\begin{frontmatter}
\title{Self consistent propagation of \\ hyperons and antikaons in nuclear matter \\based on 
relativistic chiral SU(3) dynamics}
\author[GSI]{M.F.M. Lutz}
\author[PTE]{and C.L. Korpa}
\address[GSI]{Gesellschaft f\"ur Schwerionenforschung (GSI),\\
Planck Str. 1, 64291 Darmstadt, Germany}
\address[PTE]{Department of Theoretical Physics, University of
Pecs, \\Ifjusag u.\ 6, 7624 Pecs, Hungary}
\begin{abstract}
We evaluate the antikaon spectral density in isospin symmetric nuclear matter. 
The in-medium antikaon-nucleon scattering process and the antikaon propagation
is treated in a self consistent and relativistic manner where a maximally scheme-independent 
formulation is derived by performing a partial density resummation in terms of the 
free-space antikaon-nucleon scattering amplitudes. The latter amplitudes are taken from a 
covariant and chiral coupled-channel SU(3) approach which includes s-, p- and
d-waves systematically. Particular care is taken on the proper evaluation of the in-medium
mixing of the partial waves. Our analysis establishes a rich structure of the antikaon spectral 
function with considerable strength at small energies. At nuclear saturation density we predict 
attractive mass shifts for the $\Lambda(1405)$, $\Sigma (1385)$ and $\Lambda(1520)$ of about 60 MeV, 
60 MeV and 100 MeV respectively. The hyperon states are found to exhibit at the same time an increased 
decay width of about 120 MeV for the s-wave $\Lambda(1405)$, 70 MeV for the p-wave $\Sigma (1385)$ 
and 90 MeV for the d-wave $\Lambda(1520)$ resonance.
\end{abstract}
\end{frontmatter}


\section{Introduction}

A good understanding of the antikaon spectral function in nuclear matter is required for the 
description of $K^-$-atoms \cite{Gal,Florkowski} and the subthreshold production of kaons in heavy ion 
reactions \cite{Senger}. An exciting consequence of a significantly reduced effective 
$K^-$ mass could be that kaons condense in the interior of neutron stars \cite{K:condensation:1,K:condensation:2}. 
The ultimate goal is to relate the in-medium spectral function of 
kaons with the anticipated chiral symmetry restoration at high baryon density. 
To unravel quantitative constraints on the kaon spectral functions from subthreshold kaon production data 
of heavy-ion reactions requires transport model calculations which are performed 
extensively by various groups \cite{Fuchs,Aichelin,Bratkovskaya,Ko}. The next generation of 
transport codes which are able to incorporate  more consistently particles with finite width  
are being developed \cite{Knoll,Leupold,Cassing,Schaffner}. This is of considerable 
importance when dealing with antikaons which are poorly described by a quasi-particle 
ansatz \cite{ml-sp,ramossp}.

There has been much theoretical effort to access the properties of kaons in nuclear 
matter \cite{Lutz,Pethick,Kolomeitsev,Waas1,Waas2,ml-sp,ramossp,Tolos}. An  
antikaon spectral function with support at energies smaller than the free-space kaon mass 
was already anticipated in the 70's by the many K-matrix analyses of the antikaon-nucleon 
scattering process (see e.g. \cite{Martin}) which predicted 
considerable attraction in the subthreshold scattering amplitudes. This leads in 
conjunction with the low-density theorem \cite{Dover,Lutz} to an attractive antikaon 
spectral function in nuclear matter. Nevertheless, the quantitative evaluation of the  
antikaon spectral function is still an outstanding problem. The challenge is first to develop 
an improved understanding of the vacuum antikaon-nucleon scattering process, in particular 
reliable subthreshold antikaon-nucleon scattering amplitudes are required, and secondly, to  
evaluate the antikaon spectral function in an improved many-body treatment. 

The antikaon-nucleon scattering is complicated due to the open inelastic 
$\pi \Sigma$ and $\pi \Lambda $ channels and the presence of the s-wave $\Lambda(1405)$ 
and p-wave $\Sigma(1385)$ resonances just below and the d-wave $\Lambda(1520)$ resonance 
not too far above the antikaon-nucleon threshold. Since the low-energy data on 
antikaon-nucleon scattering are not too precise the required subthreshold extrapolation 
is favorably performed imposing constraints from causality, covariance and chiral symmetry.
First intriguing works in this direction can be found in \cite{Kaiser,Oset} where 
the low-energy antikaon-proton scattering data were successfully described in terms 
of s-wave coupled SU(3) channels. These works are based on a cutoff regularized 
Lippmann-Schwinger equation where the potential is matched phenomenologically with the 
leading tree-level terms of the chiral Lagrangian. The fact that the subthreshold amplitudes 
of \cite{Kaiser} and \cite{Oset} differ significantly triggered the recent work \cite{LK} 
which systematically considers s-, p-, and d-waves. The $\chi$-BS(3) approach of \cite{LK} constitutes a 
considerably improved chiral scheme more consistent with covariance, crossing symmetry and the chiral 
counting concept. The amplitudes obtained in that scheme are particularly well suited for an application 
to the nuclear kaon dynamics, because it was demonstrated that they are approximately crossing symmetric 
in the sense that the $K N$ and $\bar K N$ amplitudes smoothly match at subthreshold energies. Therefore we 
believe that those amplitudes, which are of central importance for the nuclear kaon dynamics, will lead to 
rather reliable results for the propagation properties of kaons in dense nuclear matter. 

As was pointed out in \cite{ml-sp} the realistic evaluation of the antikaon self energy 
in nuclear matter requires a self consistent scheme. In particular the feedback effect of 
an in-medium modified antikaon spectral function on the antikaon-nucleon scattering process was 
found to be important for the $\Lambda(1405)$ resonance structure in nuclear matter. 
In this work we apply the newly developed $\chi$-BS(3) approach of \cite{LK} to antikaon propagation 
in nuclear matter in a self consistent manner. We develop a novel, maximally scheme-independent
and covariant framework in which self consistency is implemented in terms of the vacuum 
meson-nucleon scattering amplitudes. Our scheme considers for the first time the in-medium 
mixing of s-, p- and d-waves. Besides deriving a realistic antikaon spectral function we 
obtain the first quantitative results on the in-medium structure 
of the p-wave $\Sigma (1385)$ and the d-wave $\Lambda(1520)$ resonances.

\section{Self consistent nuclear antikaon dynamics}

In this section we introduce the self consistent and relativistic many-body framework 
required for the evaluation of the antikaon propagation in nuclear matter. 
First we recall the vacuum on-shell antikaon-nucleon scattering amplitude
\begin{eqnarray}
\langle \bar K^{j}(\bar q)\,N(\bar p)|\,T\,| \bar K^{i}(q)\,N(p) \rangle
&=&(2\pi)^4\,\delta^4(q+p-\bar q-\bar p )\,
\nonumber\\
&& \!\!\!\!\!\times \,\bar u(\bar p)\,
T^{ij}_{\bar K N \rightarrow \bar K N}(\bar q,\bar p ; q,p)\,u(p) \,,
\label{on-shell-scattering}
\end{eqnarray}
where  $\delta^4(..)$ guarantees energy-momentum conservation and $u(p)$ is the
nucleon isospin-doublet spinor. Note also $\bar K =(K^-, \bar K^0)$.
The vacuum scattering amplitude is decomposed into its isospin channels
\begin{eqnarray}
&&T^{i j}_{\bar K N \to \bar K N}(\bar q,\bar p \,; q,p) 
= T^{(0)}_{\bar K N}(\bar k,k;w)\,P^{ij}_{(I=0)}+
T^{(1)}_{\bar K N}(\bar k,k;w)\,P^{ij}_{(I=1)}\;,
\nonumber\\
&& P^{ij}_{(I=0)}= \frac{1}{4}\,\Big( \delta^{ij}\,1
+ \big( \vec \tau\,\big)^{ij}\,\vec \tau \,\Big)\, , \quad 
P^{ij}_{(I=1)}= \frac{1}{4}\,\Big( 3\,\delta^{ij}\,1- 
\big(\vec \tau \,\big)^{ij}\,\vec \tau \,\Big)\;,
\label{}
\end{eqnarray}
where $q, p, \bar q, \bar p$ are the initial and final kaon and nucleon 4-momenta and
\begin{eqnarray}
w = p+q = \bar p+\bar q\,,
\quad k= \half\,(p-q)\,,\quad
\bar k =\half\,(\bar p-\bar q)\,.
\label{def-moment}
\end{eqnarray}
In quantum field theory the scattering amplitudes $T^{(I)}_{\bar K N}$ follow as the 
solution of the Bethe-Salpeter matrix equation 
\begin{eqnarray}
T(\bar k ,k ;w ) &=& K(\bar k ,k ;w )
+\int \frac{d^4l}{(2\pi)^4}\,K(\bar k , l;w )\, G(l;w)\,T(l,k;w )\;,
\nonumber\\
G(l;w)&=&-i\,S_N({\textstyle
{1\over 2}}\,w+l)\,D_{\bar K}({\textstyle {1\over 2}}\,w-l) \,,
\label{BS-eq}
\end{eqnarray}
in terms of the Bethe-Salpeter kernel $K(\bar k,k;w)$, the free space nucleon 
propagator $S_N(p)=1/(\pslash-m_N+i\,\epsilon)$ and 
kaon propagator $D_K(q)=1/(q^2-m_K^2+i\,\epsilon)$. Following \cite{LK} we neglect 
here self energy corrections in the nucleon and kaon propagators. In a chiral scheme such 
effects are of subleading order. The Bethe-Salpeter equation (\ref{BS-eq}) implements 
properly  Lorentz invariance and unitarity for the two-body scattering process. The 
generalization of (\ref{BS-eq}) 
to a coupled-channel system is straightforward. 

The antikaon-nucleon scattering process is readily generalized from the 
vacuum to the nuclear matter case. In compact notation we write
\begin{eqnarray}
&& {\mathcal T} = {\mathcal K} + {\mathcal K} \cdot {\mathcal G} \cdot {\mathcal T}  \;,\quad 
{\mathcal T} = {\mathcal T}(\bar k,k; w,u) \;, \quad {\mathcal G} = {\mathcal G} (l;w,u) \,,
\label{hatt}
\end{eqnarray}
where the  in-medium scattering amplitude ${\mathcal T}(\bar k,k;w,u)$ and the two-particle 
propagator ${\mathcal G}(l;w,u)$ depend now on the 4-velocity $u_\mu$ 
characterizing the nuclear matter frame. For nuclear matter moving with a velocity 
$\vec v$ one has
\begin{eqnarray}
u_\mu =\left(\frac{1}{\sqrt{1-\vec v\,^2/c^2}},\frac{\vec v/c}{\sqrt{1-\vec v\,^2/c^2}}\right)
\;, \quad u^2 =1\,.
\label{}
\end{eqnarray}
We emphasize that (\ref{hatt}) is properly defined from a Feynman diagrammatic point of view even in the case 
where the in-medium scattering process is no longer well defined due to a broad antikaon spectral function. 
In this work we do not consider medium modifications of the interaction kernel, i.e. we 
approximate ${\mathcal K} = K$. We exclusively study the effect of a in-medium modified two-particle 
propagator ${\mathcal G}$
\begin{eqnarray}
&& \!\!\!\!\Delta S_N (p,u) = 2\,\pi\,i\,\Theta \Big(p\cdot u \Big)\,
\delta(p^2-m_N^2)\,\big( \pslash +m_N \big)\,
\Theta \big(k_F^2+m_N^2-(u\cdot p)^2\big)\,,
\nonumber\\
&&\!\!\!\!{\mathcal S}_N(p,u) = S_N(p)+ \Delta S_N(p,u)\,, \quad 
{\mathcal D}_{\bar K}(q,u)=\frac{1}{q^2-m_K^2-\Pi_{\bar K}(q,u)} \;,
\nonumber\\
&& \!\!\!\!{\mathcal G}(l;w,u) = -i\,{\mathcal S}_N({\textstyle
{1\over 2}}\,w+l,u)\,{\mathcal D}_{\bar K}({\textstyle {1\over 2}}\,w-l,u)  \;,
\label{hatg}
\end{eqnarray}
where the Fermi momentum $k_F$ parameterizes the nucleon density $\rho$ with
\begin{eqnarray}
\rho = -2\,\tr \,\gamma_0\,\int \frac{d^4p}{(2\pi)^4}\,i\,\Delta S_N(p,u) 
= \frac{2\,k_F^3}{3\,\pi^2\,\sqrt{1-\vec v\,^2/c^2}}  \;.
\label{rho-u}
\end{eqnarray}
In the rest frame of the bulk with $u_\mu=(1,\vec 0\,)$ one recovers with (\ref{rho-u}) the 
standard result $\rho = 2\,k_F^3/(3\,\pi^2)$. In this work we also refrain 
from including nucleonic correlation effects. Such effects have been estimated to be small 
at moderate densities \cite{Waas2} but should nevertheless be subject of a more complete 
consideration. The antikaon self energy $\Pi_{\bar K}(q,u)$ is evaluated self 
consistently in terms of the in-medium scattering amplitudes 
${\mathcal T}_{\bar K N}^{(I)}(\bar k,k;w,u)$
\begin{eqnarray}
\Pi_{\bar K}(q,u) &=& 2\,\tr \int \frac{d^4p}{(2\pi)^4}\,i\,\Delta S_N(p,u)\,
\bar {\mathcal T}_{\bar K N}\big({\textstyle{1\over 2}}\,(p-q),
{\textstyle{1\over 2}}\,(p-q);p+q,u \big)\,,
\nonumber\\
\bar {\mathcal T}_{\bar K N}&=& \frac{1}{4}\,{\mathcal T}_{\bar K N}^{(I=0)}+
\frac{3}{4}\,{\mathcal T}_{\bar K N}^{(I=1)} \;.
\label{k-self}
\end{eqnarray}
In order to solve the self consistent set of equations (\ref{hatt},\ref{hatg},\ref{k-self})
it is convenient to rewrite the scattering amplitude as follows
\begin{eqnarray}
&&{\mathcal T}= K+K\cdot {\mathcal G}\cdot {\mathcal T} 
= T+T\cdot \Delta {\mathcal G} \cdot {\mathcal T}\;,\quad
\Delta {\mathcal G}={\mathcal G}-G\;,
\label{rewrite}
\end{eqnarray} 
where $T=K+K\cdot G\cdot T$  is the vacuum scattering amplitude. The idea is to 
start with a set of tabulated coupled-channel scattering amplitudes $T$ in terms of which 
self consistency is implemented. Thus the vacuum interaction kernel $K$ need not to be 
specified. 

Coupled channel effects, which are known to be important 
for $\bar K N$-scattering, are included by assigning ${\mathcal T}$, ${\mathcal G}$ and $K$  
the appropriate matrix structures. Here we use the convention of \cite{LK}. For 
example the isospin zero loop matrix $\Delta {\mathcal G}^{(I=0)}$ reads
\begin{eqnarray}
\Delta {\mathcal G}^{(I=0)} = 
\left( 
\begin{array}{cccc}
\Delta {\mathcal G}_{\bar K N} & 0 & 0 & 0 \\
0& \Delta {\mathcal G}_{\pi \Sigma} & 0 & 0 \\
0 & 0 &\Delta {\mathcal G}_{\pi \Lambda} & 0 \\
0 & 0 & 0 & \Delta {\mathcal G}_{K \Xi } 
\end{array}
\right) \;.
\label{}
\end{eqnarray}
In this work we consider exclusively the effect of the in-medium modified $\bar K N$ channel.
Ultimately it would be desirable to include also the effects of the in-medium modified 
$\pi \Sigma$, $\pi \Lambda$ and $K \Xi$ channels. Since this requires a realistic 
in-medium pion propagator to be determined also by a self consistent scheme \cite{Korpa},  
we will consider such effects in a separate work. Here we approximate 
$\Delta {\mathcal G}_{\pi \Sigma} =0$ and $\Delta {\mathcal G}_{\pi \Lambda}=0$. Since the 
$K^+$ spectral density is affected only moderately by a nuclear environment 
we  assume $\Delta {\mathcal G}_{K \Xi }=0$ also.  With these assumptions 
the coupled-channel problem reduces to a single channel problem if 
rewritten in terms of the vacuum $\bar K N$ amplitude $T_{\bar K N \to \bar K N}$:
\begin{eqnarray}
&&{\mathcal T}_{\bar K N \to \bar K N}^{(I)}=  T_{\bar K N \to \bar K N}^{(I)}
+T_{\bar KN \to \bar K N}^{(I)}\cdot \Delta {\mathcal G}_{\bar KN} 
\cdot {\mathcal T}_{\bar K N \to \bar K N}^{(I)}\;,\quad
\nonumber\\
&& \Delta {\mathcal G}_{\bar K N}={\mathcal G}_{\bar K N}-G_{\bar K N}\;.
\label{rewrite:b}
\end{eqnarray} 
We point out that the self consistent set of equations (\ref{hatg},\ref{k-self},\ref{rewrite:b}) 
is now completely determined by the vacuum amplitudes $T_{\bar K N \to \bar K N}^{(I)}$. 
Note that even though our antikaon spectral function is a functional of the elastic $\bar K N \to \bar K N$ 
scattering process only, the inelastic channels $\bar K N \to X$ are nevertheless affected by 
$\Delta {\mathcal G}_{\bar K N}$. We find:
\begin{eqnarray}
&&{\mathcal T}^{(I)}_{\bar K N \to X} = T^{(I)}_{\bar K N \to X} + 
{\mathcal T}^{(I)}_{\bar K N \to \bar K N} \cdot \Delta {\mathcal G}_{\bar K N} 
\cdot T^{(I)}_{\bar K N \to X} \;,
\nonumber\\
&& {\mathcal T}^{(I)}_{Y \to X} = T^{(I)}_{Y \to X}+ 
T^{(I)}_{Y \to \bar K N} \cdot \Delta {\mathcal G}_{\bar K N} \cdot 
{\mathcal T}^{(I)}_{\bar K N \to X}\;,
\label{inelastic}
\end{eqnarray}
where $X, Y \neq \bar K N$. In (\ref{inelastic}) we derived the effect of 
$\Delta {\mathcal G}_{\bar K N}$ on the reaction $X \to Y$ with $X,Y\neq \bar K N$.

In the following section we will outline how to solve the self consistent set of 
equations (\ref{hatg},\ref{k-self},\ref{rewrite:b}) by introducing an appropriate 
projector algebra. Our scheme takes into account the medium induced mixing of partial waves 
usually neglected in conventional G-matrix calculations (see e.g. \cite{Tolos}).

\section{Projector algebra method in nuclear matter}

We proceed by briefly recalling the coupled-channel theory of \cite{LK}. 
The scattering amplitudes were obtained in a relativistic chiral $SU(3)$ approach 
which includes s-, p- and d-waves systematically. The parameters were successfully 
adjusted to reproduce the low-energy pion-nucleon and $K^\pm$-proton scattering data. 
The scattering amplitudes were decomposed  systematically into covariant projectors $Y^{(\pm)}_n$ with 
good angular momentum $J=n+\half$: 
\begin{eqnarray}
T^{(I)}(\bar k,k;w) &=& \sum_{n=0}^\infty\,
Y^{(+)}_n (\bar q, q;w)\,M^{(I,+)}(\sqrt{s}\,,n)
\nonumber\\
&+& \sum_{n=0}^\infty\,
 Y^{(-)}_n (\bar q, q;w)\,M^{(I,-)}(\sqrt{s}\,,n) 
\label{t-vacuum}
\end{eqnarray}
where $w^2=s$ and $k= \half\,(p-q)$ and $\bar k =\half\,(\bar p-\bar q)$. 
The representation (\ref{t-vacuum}) 
implies a particular off-shell behavior of the scattering amplitude, which was optimized as to simplify the 
task of solving the covariant Bethe-Salpeter equation. This is legitimate because the off-shell structure of the 
scattering amplitude is not determined by the two-body scattering processes in any case and moreover reflects the 
choice of meson and baryon interpolation fields only \cite{off-shell,Fearing}. 
The leading projectors relevant for the $J= {\textstyle{1\over 2}}$ 
and  $J= {\textstyle{3\over 2}}$ channels are $ Y_0^{(+)}$ (s-wave), $ Y_0^{(-)}$ (p-wave) 
and $ Y_1^{(+)}$ (p-wave), $ Y_1^{(-)}$ (d-wave) where:
\begin{eqnarray}
Y_0^{(\pm )}(\bar q,q;w) &=& \frac{1}{2}\,\left( \frac{\wslash}{\sqrt{w^2}}\pm 1 \right)\;,
\nonumber\\
Y_1^{(\pm)}(\bar q,q;w) &=& 
\frac{3}{2}\,\left( \frac{\wslash}{\sqrt{w^2}}\pm 1 \right)
\left(\frac{(\bar q \cdot w )\,(w \cdot q)}{w^2} -\big( \bar q\cdot q\big)\right)
\nonumber\\
&- &\frac{1}{2}\,\Bigg(  \barqslash -\frac{w\cdot \bar q}{w^2}\,\wslash \Bigg)
\Bigg(\frac{\wslash}{\sqrt{w^2}}\mp 1\Bigg)\,
\Bigg( \qslash -\frac{w\cdot q}{w^2}\,\wslash \Bigg)\;.
\label{}
\end{eqnarray}
For instance the projector $Y_1^{(-)}$ probes the d-wave $\Lambda (1520)$ 
resonance and $Y_1^{(+)}$ the p-wave $\Sigma (1385)$ resonance. 
For more details on the construction and the properties of these projectors we refer to \cite{LK}. 

In the following we will solve the self consistent set of equations 
(\ref{rewrite:b},\ref{hatg},\ref{k-self}) with $n=0,1$ in (\ref{t-vacuum}). The solution 
of the Bethe-Salpeter equation in nuclear matter is considerably complicated by the 
fact that the partial waves, which decouple in the vacuum, start to mix in the 
medium. This manifests itself in the presence of further tensor structures in the 
scattering amplitude not required in the vacuum. For example, if one starts to iterate
the Bethe-Salpeter equation (\ref{rewrite:b}) in terms of the vacuum amplitudes 
$M^{(I,\pm)}(\sqrt{s}\,, n)$ incorporating $\Delta {\mathcal G}_{\bar K N}$ as determined by (\ref{k-self}) 
with ${\mathcal T}_{\bar K N} = T_{\bar K N}$ one finds that ${\mathcal T}_{\bar K N}$ can 
no longer be decomposed into the projectors $ Y^{(\pm)}_n (\bar q, q;w)$. The 
in-medium solution of the Bethe-Salpeter equation requires a more general ansatz 
for the scattering amplitude. 

We found that the following ansatz solves the self consistent set of equations:
\begin{eqnarray}
&&{\mathcal T}(\bar k,k;w,u) = {\mathcal T}(w,u)+{\mathcal T}^\nu \,(w,u)\,q_\nu 
+\bar q_\mu \,{\bar {\mathcal T}}^\mu (w,u)+ \bar q_\mu\,{\mathcal T}^{\mu \nu}(w,u)\,q_\nu \;,
\nonumber\\ \nonumber\\
&&\bar {\mathcal T}^\mu =\sum_{j=3}^8\,\Big( {\mathcal M}_{[1j]}^{(p)}\,\bar P_{[1j]}^\mu  
+ {\mathcal M}_{[2 j]}^{(p)}\,\bar P_{[2j]}^\mu\Big)\,, \;
{\mathcal T}^\mu = \sum_{j=3}^8\,\Big( {\mathcal M}_{[j1]}^{(p)}\,P_{[j1]}^\mu  
+ {\mathcal M}_{[j2]}^{(p)}\,P_{[j2]}^\mu\Big) \,,
\nonumber\\
&& {\mathcal T} = \sum_{i,j=1}^2 \,{\mathcal M}^{(p)}_{[ij]}\,P_{[ij]} \,, \quad 
{\mathcal T}^{\mu \nu} = \sum_{i,j=3}^8\,{\mathcal M}_{[ij]}^{(p)}\,P_{[ij]}^{\mu \nu}+
\sum_{i,j=1}^2\,{\mathcal M}^{(q)}_{[ij]}\,Q_{[ij]}^{\mu \nu}\,,
\label{ansatz}
\end{eqnarray}
where we introduced appropriate projectors $P_{[ij]}(w,u)$ which generalize the 
vacuum projectors $ Y_n^{(\pm)}(\bar q,q;w)$. The matrix valued  functions 
${\mathcal M}_{[ij]}^{(p,q)}(w,u)$ are scalar and therefore depend only  on $s =w^2$ and 
$w \cdot u$. The set of projectors is constructed 
to satisfy the convenient algebra:
\begin{eqnarray}
&&P_{[ik]}\cdot P_{[lj]} =\delta_{kl}\,P_{[ij]} \;, \quad 
P^\mu_{[ik]}\;\bar P^\nu_{[lj]}= \delta_{kl}\,P_{[ij]}^{\mu \nu}\,,\quad 
\bar P^\mu_{[ik]}\,g_{\mu \nu}\,P^\nu_{[lj]}= \delta_{kl}\,P_{[ij]} \;,
\nonumber\\
&& Q_{[ik]}^{\mu \alpha }\,g_{\alpha \beta}\,P_{[lj]}^{\beta \nu }
= 0 = P_{[ik]}^{\mu \alpha }\,g_{\alpha \beta}\,Q_{[lj]}^{\beta \nu }
\;, \quad  
Q_{[ik]}^{\mu \alpha }\,g_{\alpha \beta}\,P_{[lj]}^{\beta }
= 0 = \bar P_{[ik]}^{\alpha }\,g_{\alpha \beta}\,Q_{[lj]}^{\beta \nu }\;,
\nonumber\\
&& Q_{[ik]}^{\mu \alpha }\,g_{\alpha \beta}\,Q_{[lj]}^{\beta \nu }
= \delta_{kl}\,Q_{[ij]}^{\mu \nu} \;,\quad 
P_{[ik]}^{\mu \alpha }\,g_{\alpha \beta}\,P_{[lj]}^{\beta \nu }
= \delta_{kl}\,P_{[ij]}^{\mu \nu}\;.
\label{proj-algebra}
\end{eqnarray}
The projector algebra (\ref{proj-algebra}) properly implements the coupling of 
the partial waves in the medium. Consider for example the projectors $P_{[ij]}(w,u)$ for 
$i,j=1,2$,
\begin{eqnarray}
&&P_{[11]}= +Y_0^{(+)} \;, \quad \, 
P_{[12]} = -Y_0^{(+)} \,
\frac{\gamma \cdot u}{\sqrt{1-(w\cdot u)^2/w^2}}\,Y_0^{(-)}\,,
\nonumber\\
&& P_{[22]}= -Y_0^{(-)} \,,\quad \,
P_{[21]} = -Y_0^{(-)} \,
\frac{\gamma \cdot u}{\sqrt{1-(w\cdot u)^2/w^2}}\,Y_0^{(+)} \,,
\label{res-proj}
\end{eqnarray}
which demonstrate that the vacuum projectors start to mix in the medium due 
to the presence of the matter 4-velocity $u_\mu$. If we neglected the $J={\textstyle{3\over 2}}$ 
amplitudes $M_1^{(\pm )}$  the Bethe-Salpeter equation 
is solved with the restricted set of projectors $P_{[ij]}$ given in (\ref{res-proj}). 
Explicit results for the remaining projectors are presented in Appendix A. In particular
we find
\begin{eqnarray}
&& Y_1^{(+)} =-3\, \bar q_\mu\, 
\Big( Q_{[11]}^{\mu \nu}+ P_{[77]}^{\mu \nu}\Big)\, q_\nu \,,\quad 
Y_1^{(-)} =+3\, \bar q_\mu\, 
\Big( Q_{[22]}^{\mu \nu}+ P_{[88]}^{\mu \nu}\Big)\, q_\nu \;.
\label{}
\end{eqnarray}
We conclude that in the space of $P_{[ij]}$-projectors all considered partial waves  
couple whereas in the $Q_{[ij]}$-projector space only the two $J={\textstyle{3\over 2}}$ 
waves couple. This reflects the fact that the four polarizations of a spin 
${\textstyle{3\over 2}}$ fermion are no longer degenerate in a nuclear environment. In 
analogy to a spin 1 boson with one longitudinal and two degenerate transverse modes one expects two independent 
modes for a spin ${\textstyle{3\over 2}}$ fermion in nuclear matter also. We refer to these modes as $P$-space 
and $Q$-space states.

With the ansatz (\ref{ansatz}) the antikaon-nucleon propagator 
$\Delta {\mathcal  G}_{\bar K N}$ translates into the set of loop functions
\begin{eqnarray}
&& \int \frac{d^4l }{(2\pi)^4}\,
\Delta {\mathcal G}(l-\half w;w,u)
=\sum_{i,j=1}^2 \,\Delta J^{(p)}_{[ij]}(w,u)\,P_{[ij]}(w,u) \;,
\nonumber\\
&& \int \frac{d^4l }{(2\pi)^4}\,l^\mu 
\,\Delta {\mathcal G}(l-\half w;w,u)
\nonumber\\
&& \qquad \qquad \quad 
=\sum_{j=3}^8\,\Big( \Delta J_{[j1]}^{(p)}(w,u)\,P_{[j1]}^\mu (w,u) 
+ \Delta J_{[j2]}^{(p)}(w,u)\,P_{[j2]}^\mu(w,u)\Big)
\nonumber\\
&& \qquad \qquad \quad 
= \sum_{j=3}^8\,\Big( \Delta J_{[1j]}^{(p)}(w,u)\,\bar P_{[1j]}^\mu(w,u)  
+ \Delta J_{[2j]}^{(p)}(w,u)\,\bar P_{[2j]}^\mu(w,u)\Big)\;,
\nonumber\\
&& \int \frac{d^4l }{(2\pi)^4}\,l^\mu \,l^\nu \,
\Delta {\mathcal G}(l-\half w;w,u)
\nonumber\\
&& \qquad \qquad \quad 
= \sum_{i,j=3}^8\,\Delta J_{[ij]}^{(p)}(w,u)\,P_{[ij]}^{\mu \nu}(w,u)
+\sum_{i,j=1}^2\,\Delta J^{(q)}_{[ij]}(w,u)\,Q_{[ij]}^{\mu \nu}(w,u)  \;,
\nonumber\\
&& \Delta {\mathcal G}(l-\half w;w,u) =-i\,\Big(
{\mathcal S}_N (l,u)\,{\mathcal D}_K(w -l,u)
-S_N(l)\,D_K(w-l)\Big)  \,,
\label{}
\end{eqnarray}
which can be decomposed in terms of our projectors. The reduced loop functions 
$\Delta J_{[ij]}(w,u)$ acquire the generic form
\begin{eqnarray}
&&\Delta J_{[ij]}(w,u) = \int \frac{d^4l}{(2 \pi)^4}\,g(l;w,u)\,\Delta J_{[ij]}(l;w,u)\,,
\nonumber\\ \nonumber\\
&& g(l;w,u) = 2\,\pi\,\Theta \Big(l\cdot u \Big)\,\delta(l^2-m_N^2)\,
\frac{\Theta \Big(k_F^2+m_N^2-(u \cdot l)^2 \Big)}{(w-l)^2-m_K^2-\Pi_{\bar K}(w-l,u)}
\nonumber\\
&& \qquad \qquad -\frac{i}{l^2-m_N^2+i\,\epsilon} \,
\frac{1}{(w-l)^2-m_K^2-\Pi_{\bar K}(w-l,u)}
\nonumber\\
&& \qquad \qquad +\frac{i}{l^2-m_N^2+i\,\epsilon} \,\frac{1}{(w-l)^2-m_K^2+i\,\epsilon}\;,
\label{j-exp}
\end{eqnarray}
where the scalar polynomials $\Delta J_{[ij]}(l;w,u)$ involve typically some 
powers of $l^2,l \cdot w$ or $l \cdot u$. They are listed in Appendix B. We observe  
that the reduced loop functions $\Delta J_{[ij]}(w,u)$ are scalar and therefore 
depend only on $w^2$ and $w \cdot u$. Thus the loop functions can be evaluated in any 
convenient frame without loss of information. In practice we perform the loop integration 
in the rest frame of nuclear matter with $u_\mu=(1,\vec 0)$. It is straightforward 
to perform the energy and azimuthal angle integration in (\ref{j-exp}). The energy integration 
of the last two terms in (\ref{j-exp}) is performed by closing the complex contour in the 
lower complex half plane. One picks up two contributions: first the nucleon pole leading 
to $l_0=\sqrt{m_N^2+\vec l^2}$ in (\ref{j-exp}) and second the kaon pole at typically 
$w_0-l_0 = -\sqrt{m_K^2+(\vec l-\vec w\,)^2}$. Here we neglect the kaon pole contribution 
since it is expected to be small due to the only moderate change of the kaon spectral function in 
nuclear matter: for the kaon pole contribution the second and third term in (\ref{j-exp}) 
cancel to good accuracy. All together one is left with a two-dimensional integral which must 
be evaluated numerically. In our numerical simulation we restrict $|\vec l\,| < 800$ MeV
as to avoid poorly controlled contributions from the antikaon-nucleon scattering amplitudes for 
energies larger than about $\sqrt{s} \simeq 1700$ MeV. 

Note that the center of mass frame of the antikaon-nucleon system with 
$w_\mu = (\sqrt{s},\vec 0\,)$ does not necessarily 
lead to any further simplification as suggested in \cite{ramossp} since in this frame a nonzero 
bulk velocity $\vec v \neq 0 $ is required. After performing the energy and azimuthal angle 
integration in (\ref{j-exp}) one is left with a two dimensional integration which must 
be evaluated numerically. The scalar self energy $\Pi_{\bar K}(l-w,u)$ depends on 
the two invariants $(l-w)^2$ and $(l-w)\cdot u$ only. Consequently in the nuclear matter rest 
frame the first entry involves the angle $\vec l \cdot \vec w$ as compared to the 
frame with $w_\mu= (\sqrt{s},\vec 0\,)$ which leads the angle $\vec l \cdot \vec v$ 
in second entry. This confirms the expected conservation of complexity. A  
strong 'vector' potential in the antikaon self energy, as suggested by phenomenology, 
actually means a strong dependence on $(l-w)\cdot u $ in (\ref{j-exp}). Therefore it 
appears unjustified to neglect that dependence as was done in \cite{ramossp,Tolos,Schaffner}.

Finally the in-medium Bethe-Salpeter equation reduces to a simple matrix equation
\begin{eqnarray}
&& {\mathcal M}^{(p)}_{[ij]}(w,u) = M^{(p)}_{[ij]}(\sqrt{s}\,) 
+ \sum_{l,k\,=1}^8\,M^{(p)}_{[ik]}(\sqrt{s}\,) \,\Delta J^{(p)}_{[kl]}(w,u) \, 
{\mathcal M}^{(p)}_{[lj]}(w,u) \;, 
\nonumber\\
&& {\mathcal M}^{(q)}_{[ij]}(w,u) = M^{(q)}_{[ij]}(\sqrt{s}\,)
+ \sum_{l,k\,=1}^2\,M^{(q)}_{[ik]}(\sqrt{s}\,) \,\Delta J^{(q)}_{[kl]}(w,u) \, 
{\mathcal M}^{(q)}_{[lj]}(w,u) \;,
\label{}
\end{eqnarray}
where $s=w_0^2-\vec w\,^2$. We identify the nonzero vacuum matrix elements $M^{(p,q)}_{[ij]}(\sqrt{s}\,)$ by
\begin{eqnarray}
&& M^{(q)}_{[11]}(\sqrt{s}\,)  = -3\,M^{(+)}(\sqrt{s}\,,1)  \, , \quad 
M^{(q)}_{[22]}(\sqrt{s}\,)  = 3\, M^{(-)} (\sqrt{s}\,,1) \,,\quad
\nonumber\\
&& M^{(p)}_{[11]}(\sqrt{s}\,)  = M^{(+)}(\sqrt{s}\,,0)  \, , \quad 
M^{(p)}_{[22]}(\sqrt{s}\,)  = - M^{(-)}(\sqrt{s}\,,0)  \,, \quad 
\nonumber\\
&& M^{(p)}_{[77]}(\sqrt{s}\,)  = -3\,M^{(+)}(\sqrt{s}\,,1)  \, , 
\quad M^{(p)}_{[88]}(\sqrt{s}\,)  = 3\, M^{(-)}(\sqrt{s}\,,1)  \,.
\label{m:eq}
\end{eqnarray}
Note that in (\ref{m:eq}) the isospin index $I$ is suppressed. We observe that 
in fact all invariant amplitudes ${\mathcal M}^{(p)}_{[ij]}(w,u)$ with either $i\in \{3,4,5,6\}$ or
$j\in \{3,4,5,6\}$ are zero. That is a direct consequence of the analogous property for the 
vacuum amplitudes $M_{[ij]}$. We emphasize that in principle the vacuum scattering amplitudes could 
have contributions proportional to $P_{[33]}, P_{[44]}, P_{[55]}$ or $P_{[66]}$ also. The reason why we do not 
consider such terms is that they are not determined by the on-shell antikaon-nucleon scattering process. 
The size of such terms merely reflects the choice of the interpolating fields \cite{off-shell,Fearing} and 
therefore were not determined in \cite{LK}. They acquire physical significance only, if 
at the same time the free-space 3-body scattering processes are evaluated systematically. 
The antikaon self energy follows
\begin{eqnarray}
\Pi_{\bar K}(q,u) &=& -\sum_{i,j=1}^8\,\int_0^{k_F} \frac{d^3 p}{(2\pi)^3} \,
\frac{2}{E_p}\,c^{(p)}_{[ij]}(q;w,u)\,\bar {\mathcal M}^{(p)}_{[ij]}(w,u) 
\nonumber\\
&-& \sum_{i,j=1}^2\,\int_0^{k_F} \frac{d^3 p}{(2\pi)^3} \,
\frac{2}{E_p}\,c^{(q)}_{[ij]}(q;w,u)\,\bar {\mathcal M}^{(q)}_{[ij]}(w,u) \,,
\label{kaon-final}
\end{eqnarray}
where we used $u_\mu=(1, \vec 0\,), q_\mu=(\omega, \vec q\,)$ and 
$w_\mu = (\omega+E_p, \vec q+\vec p\, )$ with $E_p=(m_N^2+\vec p\,^2)^{1/2}$. The scalar 
coefficient functions $c^{(p,q)}_{[ij]}(q;w,u)$ are listed in Appendix C. In (\ref{kaon-final})
we introduced the isospin averaged amplitudes:
\begin{eqnarray}
\bar {\mathcal M}_{[ij]}(w,u) = \frac{1}{4}\,{\mathcal M}^{(I=0)}_{[ij]}(w,u) 
+\frac{3}{4}\,{\mathcal M}^{(I=1)}_{[ij]}(w,u) \,.
\label{}
\end{eqnarray}
With (\ref{kaon-final}), (\ref{m:eq}) and (\ref{j-exp}) we arrive at our final 
self consistent set of equations which is solved numerically by iteration. First 
one determines the leading antikaon self energy $\Pi_{\bar K}(\omega ,\vec q\,)$ 
by (\ref{kaon-final}) with ${\mathcal M}_{[ij]} =M_{[ij]}$. That leads via the loop 
functions (\ref{j-exp}) and the in-medium Bethe-Salpeter equation (\ref{m:eq}) to medium 
modified scattering amplitudes ${\mathcal M}_{[ij]}(w_0,\vec w\,) $. The latter are used 
to determine the antikaon self energy of the next iteration. This procedure typically 
converges after 3 to 4 iterations. The manifest covariant form of the self 
energy and scattering amplitudes are recovered with 
$\Pi_{\bar K}(q^2,\omega)= \Pi_{\bar K}(q^2,q \cdot u)$ 
and ${\mathcal M}_{[ij]}(w^2,w_0)={\mathcal M}_{[ij]}(w^2,w \cdot u)$ in a straightforward 
manner if considered as functions of $q^2,\omega $ and $w^2, w_0$ respectively.

\section{Results}

We developed a self consistent microscopic theory for the propagation of hyperons and antikaons in cold nuclear matter. 
The many-body formulation of the previous sections requires as input exclusively the partial-wave antikaon-nucleon 
scattering amplitudes. Unfortunately a partial wave analysis of elastic and inelastic antikaon-nucleon scattering data 
without further constraints from theory is inconclusive at present \cite{Dover,Gensini}. Comparing for instance the 
energy dependent analyses \cite{gopal} and \cite{garnjost} one finds large uncertainties in the s- and p-waves in 
particular at low energies. This reflects on the one hand a model dependence of the analysis and on the other hand 
an insufficient data set. Moreover, of central importance for the derivation of a realistic antikaon spectral function 
are the subthreshold $\bar K N$ scattering amplitudes which require further poorly constrained extrapolations. Since 
the antikaon spectral function tests the $\bar K N$ amplitudes at subthreshold energies where they are not directly 
determined by empirical data it is crucial to take $\bar K N$ amplitudes as input for the many-body calculation which 
are consistent with constraints set by causality, chiral symmetry and crossing symmetry. As was pointed out in 
\cite{Hurtado} the rather ancient analysis of \cite{Kim}, even though troubled with severe shortcomings \cite{piN:Hurtado},
was the only one which included s- and p-waves and still reproduced the most relevant features of the subthreshold 
$\bar K N$ amplitudes. We therefore believe that the amplitudes of the recently developed $\chi$-BS(3) approach, 
which systematically incorporated constraints from chiral symmetry, crossing symmetry and causality, are 
best suited for an application to the nuclear antikaon dynamics. For a detailed presentation of the 
$\chi$-BS(3) approach we refer to \cite{LK}. The amplitudes of that work, which will be recalled in part when 
presenting their in-medium modifications, describe the antikaon-nucleon scattering process quantitatively up to 
$p_{\rm lab.} \simeq 500$ MeV and qualitatively up to about $p_{\rm lab.} \simeq 1000$ MeV.

\subsection{Hyperon resonances in nuclear matter}

We give a presentation and discussion of our results for the in-medium modification of the hyperon 
resonance properties. The resonance propagator may be identified with the appropriate 
antikaon-nucleon scattering amplitude of a given partial wave. In the self consistent scheme of section 3  
the in-medium scattering process is intimately related to the antikaon spectral function, for which 
our results will be presented in the next section. According to (\ref{kaon-final}) once the 
self consistent in-medium scattering process is established the antikaon self energy follows 
by averaging the in-medium scattering amplitudes over the Fermi distribution (\ref{kaon-final}). 
Therefore the pertinent structures in the in-medium amplitudes already tell the characteristic features expected 
in the antikaon spectral function. The $\chi$-BS(3) approach of \cite{LK}, applied here to the nuclear antikaon dynamics, 
encounters many hyperon resonances. Firmly established are the s-wave 
$\Lambda (1405)$, the p-wave $\Sigma (1385)$ and the d-wave $\Lambda(1520)$ resonances, for all of which we 
derive their in-medium properties. Less reliably established are the s-wave $\Lambda(1800)$, the p-wave 
$\Lambda (1600)$ and the two d-wave $\Sigma (1690)$ and $\Lambda (1680)$ resonances 
because their properties were not directly constrained by the empirical data set. It was a highly nontrivial
but expected result of \cite{LK}, that all resonances but the p-wave baryon decuplet and the d-wave baryon nonet 
resonances were  generated dynamically by the chiral coupled-channel dynamics once agreement with the low-energy 
data set with $p_{\rm lab.}< 500$ MeV was achieved. A more accurate description of the latter resonances 
requires the extension of the $\chi$-BS(3) approach by including more inelastic channels. 

A particularly interesting phenomenon observed in our approach is the in-medium induced mixing 
of resonances with different quantum numbers $J^P$. In a given isospin channel we find that at 
vanishing three momentum of the meson-baryon state $|\vec w |= 0$ all four partial wave amplitudes 
considered in this work decouple identically. However, 
once the meson-baryon pair is moving relative to the nuclear matter bulk with $|\vec w | \neq 0$ 
we find two separate channels only. In the first channel, our P-space, all considered partial wave 
amplitudes $S_{01}$, $P_{01}$, $P_{03}$ and $D_{03}$ couple, whereas in the second channel, our Q-space, only the partial 
wave amplitudes $P_{03}$ and $D_{03}$ do. The fact that the $J={\textstyle{3\over 2}}$ amplitudes $P_{03}$ 
and $D_{03}$ affect both the P- and Q-space is not surprising, because for those states one would expect an 
in-medium splitting of their four modes which are degenerate in free-space. This is somewhat analogous to the 
longitudinal and transverse modes of vector-mesons, which also split in nuclear matter. 

\begin{figure}[t]
\begin{center}
\includegraphics[width=14cm,clip=true]{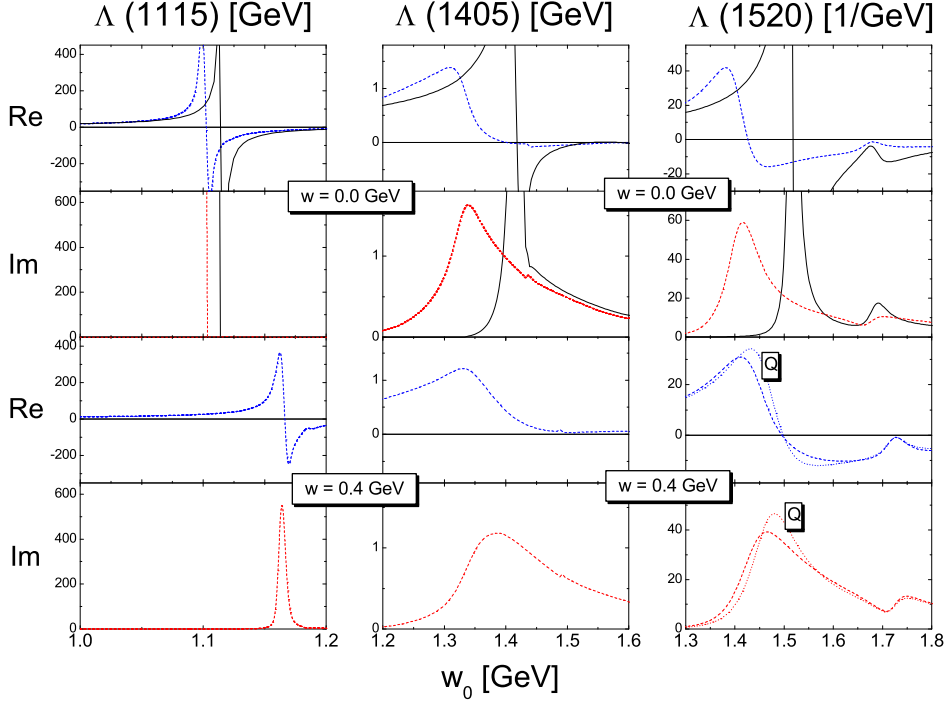}
\end{center}
\caption{In-medium properties of $\Lambda$ hyperons with energy $w_0$ and momentum $\vec w$ at nuclear saturation 
density $\rho= 0.17$ fm$^{-3}$. The amplitudes plotted with dashes lines are 
$-f^2\,M^{(p)}_{[22]}$ for the $\Lambda(1115)$, $f^2\,M^{(p)}_{[11]}$ for the $\Lambda(1405)$ and 
$f^2\,M^{(p)}_{[88]}/3,\,f^2\,M^{(q)}_{[22]}/3$ for the $\Lambda(1520)$ where $f=90$ MeV (see (\ref{m:eq})). 
The solid lines give the corresponding amplitudes in the free-space limit with $\rho= 0$.}
\label{fig:hyperons:I=0}
\end{figure}

We summarize that one expects sizeable effects from the in-medium mixing of the partial-wave 
amplitudes at $\vec w \neq 0$ if any two of the $S_{01}$, $P_{01}$, $P_{03}$ or $D_{03}$ 
amplitudes show significant strength at a given energy $w_0$ simultaneously. To some extent 
this is the case for the amplitudes reflecting the $\Sigma (1195)$ ground state and the p-wave 
$\Sigma (1385)$ resonance for which we anticipate important mixing effects at large densities. 
Similarly we would predict that a $\Lambda(1115)$ moving relative to the nuclear matter bulk may show 
an interesting interplay with the s-wave $\Lambda (1405)$ resonance.

\begin{figure}[t]
\begin{center}
\includegraphics[width=14cm,clip=true]{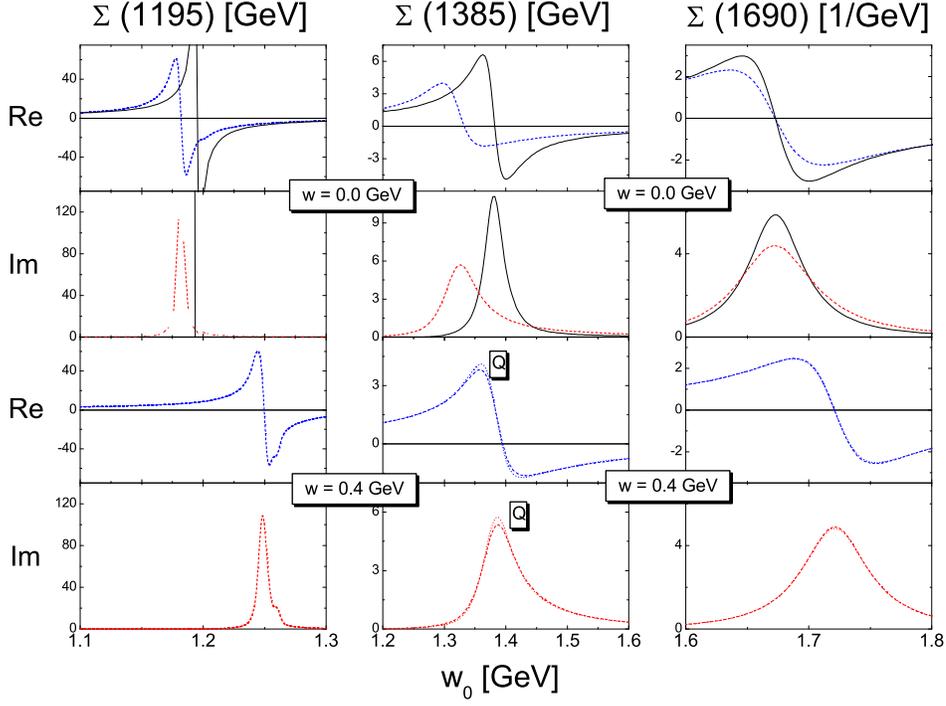}
\end{center}
\caption{In-medium properties of $\Sigma$ hyperons with energy $w_0$ and momentum $\vec w$ at 
nuclear saturation density $\rho= 0.17$ fm$^{-3}$. The amplitudes plotted with dashed lines are 
$-f^2\,M^{(p)}_{[22]}$ for the $\Sigma(1115)$, $-f^2\,M^{(p)}_{[77]}/3,\,-f^2\,M^{(q)}_{[11]}/3$ for 
the $\Sigma(1385)$ and $f^2\,M^{(p)}_{[88]}/3,\,f^2\,M^{(q)}_{[22]}/3$ for the $\Sigma(1690)$ 
where $f=90$ MeV (see (\ref{m:eq})).
The solid lines give the corresponding amplitudes in the free-space limit with $\rho= 0$.}
\label{fig:hyperons:I=1}
\end{figure}

In Fig. \ref{fig:hyperons:I=0} we present our results for the in-medium propagation of the $\Lambda(1115)$ ground 
state, the s-wave $\Lambda(1405)$ and the d-wave $\Lambda(1520)$ resonance all at nuclear saturation density
$\rho_0 = 0.17$ fm$^{-3}$. A $\Lambda (1115)$ with zero momentum $\vec w =0$ is described well by a quasi-particle 
state with an attractive energy shift of about 10 MeV and zero decay width.  However, as shown 
in Fig. \ref{fig:hyperons:I=0} at a three-momentum of $|\vec w |= 400$ MeV the $\Lambda(1115)$ receives a 
small width of about 2 MeV. That effect may be interpreted as an mixing of the  p-wave $\Lambda (1115)$ state with the 
s-wave $\Lambda(1405)$ resonance. At $|\vec w |= 0$ the s-wave resonance by itself is considerably broadened and 
subject to an attractive energy shift of about 60 MeV. Here we find a considerably stronger in-medium effect for the 
$\Lambda(1405)$ as compared to previous works \cite{ml-sp,ramossp}. That large attractive mass shift found for the 
$\Lambda (1405)$ resonance is due to p-wave effects. Putting all p- and d-wave amplitudes to zero in our scheme, a self 
consistent evaluation of the $\Lambda (1405)$ structure leads to an attractive mass shift of less than 10 MeV. 
In Fig. \ref{fig:hyperons:I=0} it is shown also a stunning medium modification for the 
d-wave $\Lambda(1520)$ resonance which acquires a width of about 90 MeV and an attractive mass shift of 
about 100 MeV. The dependence of the resonance energy on the momentum appears quite compatible with a free space 
dispersion relation $w_0 = \sqrt{m^2+\vec w\,^2}$ when interpreted in terms of the effective in-medium modified rest 
mass. The splitting of the P-space and Q-space modes for the 
d-wave resonance is found to be rather small at $|\vec w |= 400$ MeV and nuclear saturation density. 

We turn to the hyperon states with isospin one, the results for which are given in Fig. \ref{fig:hyperons:I=1}.
At nuclear saturation density the $\Sigma (1195)$ receives a width of about 2 MeV and a 
small attractive mass shift of about $10$ MeV. Its dispersion relation is quite consistent with that one known  
from free space. Interesting in-medium modifications are also observed for the  $\Sigma (1385)$ 
resonance with an attractive mass shift of about 60 MeV and an decay width increased to about 70 MeV. Again a free space 
dispersion relation predicts well the quasi-particle energy of the $\Sigma (1385)$ resonance at $|\vec w | \neq 0$. 
Moreover the P- and Q-modes of  that $J={\textstyle{3\over 2}}$ resonance are basically degenerate. The only hyperon 
resonance for which we do not predict sizeable in-medium modifications in our present scheme is the d-wave 
$\Sigma (1690)$ resonance. As seen in Fig. \ref{fig:hyperons:I=1} that resonance is very little affected by its nuclear 
environment. That may simply reflect that the $\Sigma (1690)$ resonance has a rather small branching 
fraction of about 10 $\%$ into the $\bar K N$ channel only, even though the available phase space for that channel is 
large. 

It may be worth giving an interpretation for the strong in-medium effects presented here and 
thereby anticipate part of our results for the antikaon spectral function. The medium modification of 
a given hyperon state is determined by its interaction with the nucleons. Our self consistent approach considers in 
particular the t-channel antikaon exchange contribution for that interaction. A dramatic in-medium modification of the 
antikaon spectral function as will be established below, then necessarily leads to strong effects also for the 
hyperon states. For instance if the antikaon spectral function shows strength at energies $\omega < m_\Lambda -m_N$ 
the $\Lambda (1115) \to \bar K  N$ may decay into a nucleon and an effective antikaon-like mode giving it a finite 
hadronic decay width.

\subsection{Antikaon spectral function in nuclear matter}

\begin{figure}[t]
\begin{center}
\includegraphics[width=14cm,clip=true]{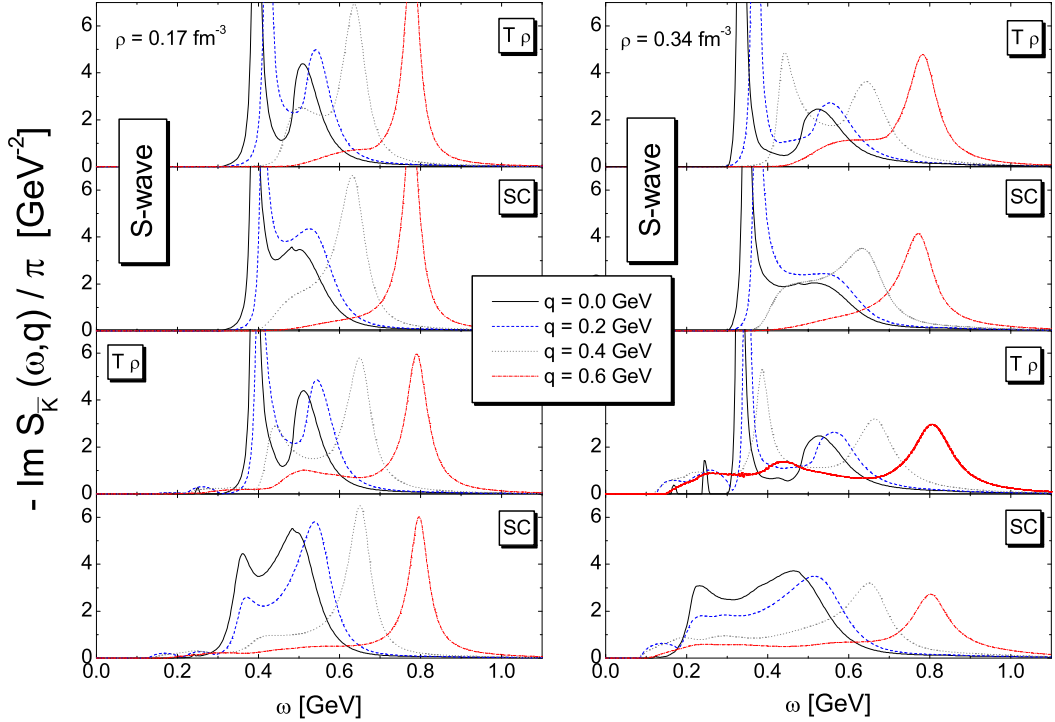}
\end{center}
\caption{Antikaon spectral function as a function of antikaon energy $\omega$ and momentum $q$.
The labels 'T$\rho$' and 'SC' refer to calculations obtained in terms of free-space and in-medium $\bar K N$ 
amplitudes respectively (see (\ref{k-self})). The first two rows give the results with only s-wave interactions  
and the last two rows with all s-, p- and d-wave contributions.}
\label{fig:kaon-sp}
\end{figure}

In Fig. \ref{fig:kaon-sp} we present the antikaon spectral function evaluated 
at nuclear densities $\rho_0$ and $2\,\rho_0$ according to various approximation strategies. 
The spectral functions exhibit a rich structure with a pronounced dependence on the antikaon three-momentum.
That reflects the presence of the various hyperon states in the $\bar K N$ amplitudes. 
Typically the peaks seen are quite broad and not always of quasi-particle type. 
In the first and third row the antikaon self energy is computed in terms of the free-space scattering amplitudes
only. Here the first row gives the result with only s-wave contributions and the third row includes all 
s-, p- and d-wave contributions established in this work. The second and fourth row give results obtained in the 
self consistent approach developed in section 3 where the full result of the last row includes all partial 
waves and the results in the second row follow with s-wave contributions only. 
We observe that in all cases a self consistent evaluation of the spectral function leads to important changes 
in the spectral function as compared to a calculation which is based on the free-space scattering amplitudes only.
Furthermore we assure that the sum rule,   
\begin{eqnarray}
-\int \frac{d \, \omega }{\pi}\,\omega \,\Im \,S_{\bar K}(\omega,\,\vec q\,) =1 \;,
\label{k-sumrule}
\end{eqnarray}
is satisfied to good accuracy. A violation of that sum rule (\ref{k-sumrule}) would indicate that 
the microscopic theory is in conflict with constraints set by causality. 

For s-wave input only the amount of attraction found in this work is similar to previous works \cite{ml-sp,ramossp}. 
As is evident upon comparing the first and third rows of Fig. \ref{fig:kaon-sp} the p- and d-wave contributions
add quite significant attraction for small energies and large momenta.  At about twice nuclear saturation density we find 
most striking the considerable support of the spectral function at small energy. That reflects the coupling of 
the antikaon to a $\Lambda(1115)$ nucleon-hole state. Upon inspecting the Lindhard function of that contribution 
for a free-space $\Lambda(1115)$ \cite{Kolomeitsev} one finds that the antikaon spectral function should be 
non-zero for
\begin{eqnarray}
\omega > \sqrt{m_\Lambda^2+(|\vec q \,|-k_F)^2}-\sqrt{m_N^2+k_F^2} \,.
\label{lam-kin}
\end{eqnarray}
The strength of the spectral function is largest at $|\vec q \,| \sim k_F$ because of the 
kinematical consideration (\ref{lam-kin}). The effects induced by the $\Lambda(1115)$ hyperon exchange 
derived here are somewhat larger than suggested in previous work \cite{Kolomeitsev}. This is due to
our attractive mass shift of about $40$ MeV for the $\Lambda (1115)$.  
The results at $2 \,\rho_0$ should be considered 
cautiously because nuclear binding and correlation effects were not yet included in the present scheme.

\section{Summary and outlook}

The microscopic $\chi$-BS(3) dynamics of \cite{LK} was applied to antikaon and hyperon resonance propagation 
in cold nuclear matter in a manifest covariant and self consistent manner. 
Of central importance for the microscopic evaluation of the antikaon spectral function in nuclear matter 
are the antikaon-nucleon scattering amplitudes, in particular at subthreshold energies. 
The required amplitudes were well established by the $\chi$-BS(3) approach and show 
sizeable contributions from p-waves not considered systematically so far \cite{Kolomeitsev,ml-sp,ramossp}. 
For the antikaon spectral function we predict a pronounced dependence on the three-momentum of the antikaon. 
The spectral function shows typically a rather wide structure invalidating a simple quasi-particle description. 
For instance at $\rho = 0.17$ fm$^{-3}$ the strength starts at the quite small energy $\omega \simeq $ 150 MeV.  
At nuclear saturation density we predict attractive 
mass shifts for the $\Lambda(1405)$, $\Sigma (1385)$ and $\Lambda(1520)$ of about 60 MeV, 60 MeV and 100 MeV 
respectively. The hyperon states are found to show at the same time an increased decay width of about 120 MeV 
for the s-wave $\Lambda(1405)$, 70 MeV for the p-wave $\Sigma (1385)$ and 90 MeV for the d-wave $\Lambda(1520)$ 
resonances. The attractive mass shifts for the $\Lambda $ hyperon ground states of about
10 MeV is smaller by about 20 MeV as compared to what one expects from phenomenological constraints set by 
hyper-nuclei \cite{Bando}. Our result for the $\Lambda (1115)$ is nevertheless interesting due to strong non-linear 
effects in density. At twice saturation density we predict an attractive mass shift of 40 MeV. The source of the 
non-linear effect is traced back to the t-channel antikaon exchange contribution dominating the hyperon-nucleon 
interaction. The strong density dependence of the antikaon spectral function leads to the non-trivial propagation 
properties of the $\Lambda (1115)$ state. If we corrected for the 
missing attraction by a term linear in the density we would predict that the $\Lambda (1115)$ experiences an attractive 
mass shift of about 80 MeV at twice saturation density. It is interesting to speculate to what extent this phenomenon 
has consequences for the formation of exotic $\Lambda$-nuclei. That may deserve further detailed studies. 
Also it would be useful to explore whether it is feasible to confirm the strong in-medium modifications of hyperon 
states suggested in this work by suitable experiments at ELSA or MAMI.  

We expect our analysis to pave the way for a microscopic description of kaonic atom data. The latter are 
known to be a rather sensitive test of the antikaon-nucleon dynamics \cite{Gal,Florkowski}. An extended domain of applicability   
for the microscopic chiral theory is foreseen once additional inelastic channels are considered systematically. 
Also an application of the covariant and self consistent many-body dynamics, developed in this work, to the 
propagation properties of pions, vector-mesons and nucleon resonances is being pursued. Here we anticipate 
interesting in-medium effects due to the mixing of the s-, p- and d-wave nucleon resonances. Before an application of 
our results to heavy-ion reactions it is necessary to extend our many-body framework to finite temperature.

{\bfseries{Acknowledgments}}

This research was supported in part by the
Hungarian Research Foundation (OTKA) grant T030855. M.F.M. Lutz acknowledges useful discussions with 
E.E. Kolomeitsev and D. Voskresenski.

\section{Appendix A: Projector algebra}

In order to construct the desired projector algebra $P_{[ij]}$ and $Q_{[ij]}$ 
of (\ref{proj-algebra}) it is useful to find appropriate building blocks which
greatly facilitate the derivation. We introduce the objects 
$P_\pm$, $U_\pm$, $V_\mu$ and $L_\mu, R_\mu$:
\begin{eqnarray}
&& P_\pm(w) = \frac{1}{2}\left( 1\pm \frac{\wslash}{\sqrt{w^2}}\right)\, ,\quad 
U_\pm (w,u)=P_\pm(w) \,\frac{i\,\gamma \cdot u}{\sqrt{(w\cdot u)^2/w^2-1}}\,P_\mp(w)\;,
\nonumber\\
&&V_\mu (w)=\frac{1}{\sqrt{3}}\,\Big( \gamma_\mu -\frac{\wslash}{w^2}\,w_\mu \Big)
\;,\quad X_\mu(w,u)= \frac{(w\cdot u)\,w_\mu-w^2\,u_\mu}{w^2\,\sqrt{(w \cdot u)^2/w^2-1}} \;,
\nonumber\\
&&R_\mu (w,u) = -\frac{1}{\sqrt{2}}\,\Big( U_+(w,u)+U_-(w,u)\Big)\,V_\mu(w)-i\,\sqrt{\frac{3}{2}}\,X_\mu(w,u)
\, , \quad 
\nonumber\\
&& L_\mu(w,u) =-\frac{1}{\sqrt{2}}\,V_\mu(w)\, \Big( U_+(w,u)+U_-(w,u)\Big) -i\,\sqrt{\frac{3}{2}}\,X_\mu(w,u) \;,
\label{def-basic}
\end{eqnarray}
which enjoy the following properties:
\begin{eqnarray}
&& P_\pm \,P_\pm = P_\pm = U_\pm \,U_\mp\, , \quad P_\pm\,P_\mp =0 = U_\pm \,U_\pm \,,
\nonumber\\
&& V \cdot L =0 = R \cdot V \,, \quad 
L \cdot V = -{\textstyle{\sqrt{8}\over 3}}\,\big( U_++U_- \big) = V \cdot R \,, \quad 
\nonumber\\
&& V \cdot V  = L \cdot R = R \cdot L =1 \,, \quad R \cdot R = L \cdot L = {\textstyle{1\over 3}}\,,
\nonumber\\
&& P_\pm \,V_\mu = V_\mu\,P_\mp \,, \quad P_\pm \,L_\mu = L_\mu\,P_\pm \,, \quad 
P_\pm \,R_\mu = R_\mu\,P_\pm \,, 
\nonumber\\
&& U_\pm \,V_\mu =-{\textstyle{1\over 3}}\,V_\mu\,U_\mp -{\textstyle{\sqrt{8}\over 3}}\,L_\mu\,P_\mp \,, \quad 
U_\pm \,L_\mu = R_\mu \,U_\pm \,, 
\nonumber\\
&& V_\mu\,U_\pm  =-{\textstyle{1\over 3}}\,U_\mp\,V_\mu -{\textstyle{\sqrt{8}\over 3}}\,R_\mu\,P_\mp \,, \quad 
U_\pm \,R_\mu = L_\mu \,U_\pm \,.
\label{basic-prop}
\end{eqnarray}
The projectors are:
\begin{eqnarray}
&&\begin{array}{llll}
P_{[11]} = P_+  \,, & P_{[12]}= U_+ \,, & P_{[21]}=U_-\,, & P_{[22]}=P_- \,, \\
P^\mu_{[31]} = V_\mu \,P_+ \,,  & P^\mu_{[32]} = V_\mu \,U_+ \;, & 
\bar P^\mu_{[13]} = P_+\,V^\mu \;,  &  \bar P^\mu_{[23]} = U_-\,V^\mu \;,  \\
P^\mu_{[41]} = V_\mu \,U_- \;,  & P^\mu_{[42]} = V_\mu \,P_- \;, &
\bar P^\mu_{[14]} = U_+\,V^\mu\;,  & \bar P^\mu_{[24]} = P_-\,V^\mu\;,  \\
P^\mu_{[51]} = \hat w_\mu \,P_+ \;,  & P^\mu_{[52]} = \hat w_\mu \,U_+ \;, &
\bar P^\mu_{[15]} = P_+\,\hat w^\mu \;,  & \bar P^\mu_{[25]} = U_-\,\hat w^\mu \;, \\
P^\mu_{[61]} = \hat w_\mu \,U_- \;,  & P^\mu_{[62]} = \hat w_\mu \,P_- \;, &
\bar P^\mu_{[16]} = U_+\,\hat w^\mu\;, & \bar P^\mu_{[26]} = P_-\,\hat w^\mu\;, \\
P^\mu_{[71]} =   L_\mu \,P_+ \;,  & P^\mu_{[72]} =   L_\mu \,U_+ \;, &
\bar P^\mu_{[17]} = P_+\,  R^\mu \;,  & \bar P^\mu_{[27]} = U_-\,  R^\mu \;, \\
P^\mu_{[81]} =   L_\mu \,U_- \;,  & P^\mu_{[82]} =   L_\mu \,P_- \;, &
\bar P^\mu_{[18]} = U_+\,  R^\mu\;, & \bar P^\mu_{[28]} = P_-\,  R^\mu\;,
\end{array}
\nonumber\\ \nonumber\\
&& P_{[i\,j]}^{\mu \nu} = P^\mu_{[i1]}\;\bar P^\nu_{[1j]} = P^\mu_{[i2]}\;\bar P^\nu_{[2j]}\,, 
\nonumber\\ \nonumber\\
&& Q_{[11]}^{\mu \nu } =\Big( g^{\mu \nu}-\hat w^\mu\,\hat w^\nu \Big)
\,P_+ - V^\mu\,P_-\,V^\nu -L^\mu\,P_+\,R^\nu \;, 
\nonumber\\
&& Q_{[22]}^{\mu \nu } =\Big( g^{\mu \nu}-\hat w^\mu\,\hat w^\nu \Big)
\,P_- - V^\mu\,P_+\,V^\nu -L^\mu\,P_-\,R^\nu \;, 
\nonumber\\
&& Q_{[12]}^{\mu \nu }  = \Big(g^{\mu \nu}-\hat w^\mu\,\hat w^\nu \Big)\,U_+ 
+{\textstyle{1\over 3}}\,V^\mu\,U_-\,V^\nu 
\nonumber\\
&&\qquad +{\textstyle{\sqrt{8}\over 3}}\,
\Big( L^\mu\,P_+\,V^\nu +V^\mu\,P_-\,R^\nu \Big) -{\textstyle{1\over 3}}\,L^\mu\,U_+\,R^\nu\;,
\nonumber\\
&&Q_{[21]}^{\mu \nu } = \Big(g^{\mu \nu}-\hat w^\mu\,\hat w^\nu\Big)\,U_- 
+{\textstyle{1\over 3}}\,V^\mu\,U_+\,V^\nu 
\nonumber\\
&&\qquad +{\textstyle{\sqrt{8}\over 3}}\,
\Big( L^\mu\,P_-\,V^\nu +V^\mu\,P_+\,R^\nu \Big)-{\textstyle{1\over 3}}\,L^\mu\,U_-\,R^\nu\;,
\label{}
\end{eqnarray}
where $\hat w_\mu = w_\mu /\sqrt{w^2} $.
By means of (\ref{basic-prop}) it is straightforward to verify (\ref{proj-algebra}).

\section{Appendix B: Evaluation of loop functions}

The loop matrix $\Delta J^{(p)}_{[ij]}(w,u)$ and $\Delta J^{(q)}_{[ij]}(w,u)$ can be expressed 
in terms of 13 independent loop functions $\Delta J_{i=0,...,12}(w,u)$
\begin{eqnarray}
&&\Delta J_i(w,u) = \int \frac{d^4l}{(2 \pi)^4}\,g(l;w,u)\,\Delta J_i(l;w,u)\,,
\label{}
\end{eqnarray}
and
\begin{eqnarray}
&&\Delta J_0(l;w,u) =1\,,\quad  \Delta J_1(l;w,u) = (l\cdot \hat w)\,,
\quad  \Delta J_2(l;w,u) = -(l\cdot X(w,u))\,,
\nonumber\\
&&\Delta J_3(l;w,u) = \frac{1}{2}\,\Big(l^2-(l \cdot \hat w)^2+(l \cdot X(w,u) )^2 \Big)\,,
\nonumber\\
&&\Delta J_4(l;w,u) =\Big(l\cdot \hat w\Big)^2\,,\quad 
\Delta J_5(l;w,u) = \Big(l\cdot X(w,u)\Big)^2\,,
\nonumber\\
&&\Delta J_6(l;w,u) = -\Big(l\cdot X(w,u)\Big)\,
\Big(l\cdot \hat w\Big)\,,\quad 
\nonumber\\
&&\Delta J_7(l;w,u) = \frac{1}{2}\,\Big(l^2-(l \cdot \hat w)^2+(l \cdot X(w,u) )^2 \Big)\,
\Big(l\cdot \hat w \Big)\,,
\nonumber\\
&&\Delta J_8(l;w,u) = -\frac{1}{2}\,
\Big(l^2-(l \cdot \hat w)^2+(l \cdot X(w,u) )^2 \Big)\,\Big(l\cdot X(w,u) \Big)\,,
\nonumber\\
&&\Delta J_9(l;w,u) = \Big(l\cdot \hat w\Big)^3\,,\quad
\Delta J_{10}(l;w,u) = -
\Big(l\cdot \hat w\Big)^2\,\Big(l\cdot X(w,u)\Big)\,,
\nonumber\\
&&\Delta J_{11}(l;w,u) = -\Big(l\cdot X(w,u)\Big)^3\,,\quad 
\Delta J_{12}(l;w,u) =\Big(l\cdot X(w,u)\Big)^2\,
\Big(l\cdot \hat w\Big)\,,
\nonumber\\ \nonumber\\
&& g(l;w,u) = 2\,\pi\,\Theta \Big(l\cdot u \Big)\,\delta(l^2-m_N^2)\,
\frac{\Theta \Big(k_F^2+m_N^2-(u \cdot l)^2 \Big)}{(l-w)^2-m_K^2-\Pi_{\bar K}(w-l,u)}
\nonumber\\
&& \qquad \qquad -\frac{i}{l^2-m_N^2+i\,\epsilon} \,
\frac{1}{(w-l)^2-m_K^2-\Pi_{\bar K}(w-l,u)}
\nonumber\\
&& \qquad \qquad +\frac{i}{l^2-m_N^2+i\,\epsilon} \,\frac{1}{(w-l)^2-m_K^2+i\,\epsilon}\;,
\label{}
\end{eqnarray}
where the 4-vector $X_\mu(w,u)$ was introduced in (\ref{def-basic}).
The matrix elements are:
\begin{eqnarray}
&&\Delta J^{(q)}_{[11]}=m_N\,\Delta J_3+\Delta J_7\;, \quad 
\Delta J^{(q)}_{[22]}=m_N\,\Delta J_3-\Delta J_7 \;, \quad 
\Delta J^{(q)}_{[12]}=-i\,\Delta J_8\,,
\nonumber\\
&&\Delta J^{(p)}_{[11]} = m_N\,\Delta J_{0}+\Delta J_{1} \;, \quad 
\Delta J^{(p)}_{[22]} = m_N\,\Delta J_{0}-\Delta J_{1} \;,\quad 
\Delta J^{(p)}_{[12]} = -i\,\Delta J_{2}\,,
\nonumber\\
&&\Delta J^{(p)}_{[13]}=\Delta J^{(p)}_{[24]}=
\frac{1}{\sqrt{3}} \,\Big( 2\,\Delta J_3-\Delta J_5 \Big)\,,\;\;\;
\Delta J^{(p)}_{[16]}=\Delta J^{(p)}_{[25]}=
-i\,\Delta J_6,\;\;\;
\nonumber\\
&&\Delta J^{(p)}_{[15]}=m_N \,\Delta J_1+\Delta J_4\;, \quad 
\Delta J^{(p)}_{[26]}=m_N \,\Delta J_1-\Delta J_4\;,
\nonumber\\
&&\Delta J^{(p)}_{[17]}=
i\,\sqrt{\frac{2}{3}}\,\big( m_N \,\Delta J_2+\Delta J_6\big)\;,\quad
\Delta J^{(p)}_{[28]}=
i\,\sqrt{\frac{2}{3}}\,\big( m_N \,\Delta J_2-\Delta J_6\big)\,,
\nonumber\\
&&\Delta J^{(p)}_{[14]}=
-\frac{i}{\sqrt{3}}(m_N \,\Delta J_2+\Delta J_6)\;,\quad
\Delta J^{(p)}_{[23]}=
-\frac{i}{\sqrt{3}}(m_N\, \Delta J_2-\Delta J_6)\;,
\nonumber\\
&&\Delta J^{(p)}_{[18]}=\Delta J^{(p)}_{[27]}=
\sqrt{\frac{2}{3}}\,\,\Big( \Delta J_3+\Delta J_5 \Big)\,,\;\;\;
\label{}
\end{eqnarray}
and
\begin{eqnarray}
&& \Delta J^{(p)}_{[33]}=\frac{1}{3}\,\Big(m_N\,\big(2\,\Delta J_3-\Delta J_5 \big)
+\Delta J_{12}-2\,\Delta J_7 \Big)\,,
\nonumber\\
&& \Delta J^{(p)}_{[44]}=\frac{1}{3}\,\Big(m_N\,\big(2\,\Delta J_3-\Delta J_5 \big)
-\Delta J_{12}+2\,\Delta J_7 \Big)\,,
\nonumber\\
&& \Delta J^{(p)}_{[55]}= m_N \,\Delta J_4+\Delta J_9\,,\quad
\Delta J^{(p)}_{[66]}= m_N \,\Delta J_4-\Delta J_9\,,
\nonumber\\
&& \Delta J^{(p)}_{[77]}=\frac{1}{3}\,\Big(m_N\,\big(\Delta J_3-2\,\Delta J_5 \big)
+\Delta J_{7}-2\,\Delta J_{12} \Big)\,,
\nonumber\\
&& \Delta J^{(p)}_{[88]}=\frac{1}{3}\,\Big(m_N\,\big(\Delta J_3-2\,\Delta J_5 \big)
-\Delta J_{7}+2\,\Delta J_{12} \Big)\,,
\nonumber\\
&& \Delta J^{(p)}_{[35]}= \Delta J^{(p)}_{[46]}=\frac{1}{\sqrt{3}}\,\Big(2\,\Delta J_7-\Delta J_{12}\Big)\,,
\nonumber\\
&&\Delta J^{(p)}_{[37]}= \Delta J^{(p)}_{[48]}=i\,\frac{\sqrt{2}}{3}\,\Big(2\,\Delta J_8-\Delta  J_{11}\Big)\,,
\nonumber\\
&&\Delta J^{(p)}_{[57]}=i\,\sqrt{\frac{2}{3}}\,\Big(m_N\,\Delta J_6+\Delta J_{10} \Big)\,,\quad
\Delta J^{(p)}_{[68]}=i\,\sqrt{\frac{2}{3}}\,\Big(m_N\,\Delta J_6-\Delta J_{10} \Big)\,,
\nonumber\\
&&\Delta J^{(p)}_{[34]}=-\frac{i}{3}\,\Big(2\,\Delta J_8-\Delta J_{11} \Big)\,,\quad 
\Delta J^{(p)}_{[78]}=\frac{i}{3}\,\Big(5\,\Delta J_8+2\,\Delta J_{11} \Big)\,,
\nonumber\\
&&\Delta J^{(p)}_{[36]}=-\frac{i}{\sqrt{3}}\,\Big(m_N\,\Delta J_6-\Delta J_{10} \Big)\,,\quad 
\Delta J^{(p)}_{[45]}=-\frac{i}{\sqrt{3}}\,\Big(m_N\,\Delta J_6+\Delta J_{10} \Big)\,,
\nonumber\\
&&\Delta J^{(p)}_{[38]}=\frac{\sqrt{2}}{3}\,\Big( m_N\,\big(\Delta J_3+\Delta J_5 \big)-\Delta J_7-\Delta J_{12}\Big)\,,
\nonumber\\
&&\Delta J^{(p)}_{[47]}=\frac{\sqrt{2}}{3}\,\Big( m_N\,\big(\Delta J_3+\Delta J_5 \big)+\Delta J_7+\Delta J_{12}\Big)\,,
\nonumber\\
&&\Delta J^{(p)}_{[58]}=\Delta J^{(p)}_{[67]}=\sqrt{\frac{2}{3}}\,\Big( \Delta J_7+\Delta J_{12} \Big)
\,,\quad \Delta J^{(p)}_{[56]}=-i\,\Delta J_{10}\,,
\label{}
\end{eqnarray}
where the remaining elements follow from the symmetry property 
$\Delta J^{(p,q)}_{[ij]}=\Delta J^{(p,q)}_{[ji]}$. We note that at $\vec w= 0$ the set of loop functions $\Delta J_n$
simplify as 
\begin{eqnarray}
\Delta J_2=\Delta J_6= \Delta J_8 =0 = \Delta J_3+\Delta J_5=\Delta J_7+\Delta J_{12} \,.
\label{}
\end{eqnarray}
That leads to the decoupling of all partial wave amplitudes at $\vec w = 0$.

\section{Appendix C: Evaluation of kaon self energy}

In the expression for antikaon self energy (\ref{k-self}) the term 
\begin{eqnarray}
&&\frac{1}{2}\,{\rm{tr}}\; [ (\FMslash p+m_N)\,{\mathcal T}_{\bar K N} (\half (p-q),\half (p-q);p+q,u)]
\nonumber\\
=&&\sum_{i,j=1}^8\,c^{(p)}_{[ij]}(q;w,u)\,{\mathcal M}^{(p)}_{[ij]}(w,u) +
\sum_{i,j=1}^2\,c^{(q)}_{[ij]}(q;w,u)\,{\mathcal M}^{(q)}_{[ij]}(w,u) \,,
\end{eqnarray}
appears with $w=p+q$. The coefficient functions $c^{(q)}_{[ij]}(q;w,u)$ and $c^{(p)}_{[ij]}(q;w,u)$ are:
\begin{eqnarray}
&&c_{[11]}^{(q)}=\frac{1}{2}\,E_+\,\Big(E_+\,E_- +(X \cdot q)^2 \Big)\,,\quad
c_{[11]}^{(p)}=E_+\,,\quad 
\nonumber\\
&&c_{[12]}^{(q)}= -\frac{i}{2}\,(X \cdot q)\,\Big( E_+\,E_-
+(X \cdot q)^2\Big)\,,\quad c_{[12]}^{(p)}=-i\,(X \cdot q)\,,
\nonumber\\
&& c_{[22]}^{(q)}=\frac{1}{2}\,E_-\,\Big(E_+\,E_- +(X \cdot q)^2 \Big) \,, \quad 
c_{[22]}^{(p)}=E_-\,, 
\nonumber\\
&&c_{[13]}^{(p)}=c_{[24]}^{(p)}= -\frac{1}{\sqrt{3}}\,E_+\,E_-\,,\quad 
c_{[25]}^{(p)}=c_{[16]}^{(p)}=-i\,(\hat w \cdot q)\,(X \cdot q)\,,
\nonumber\\
&&c_{[17]}^{(p)}= -i\,\sqrt{\frac{2}{3}}\,E_+ \,(X \cdot q)\,,\quad 
c_{[15]}^{(p)}= (\hat w \cdot q)\,E_+\,,\quad 
c_{[14]}^{(p)}= \frac{i}{\sqrt{3}}\,E_+\,(X \cdot q) \,,
\nonumber\\
&&c_{[28]}^{(p)}= -i\,\sqrt{\frac{2}{3}}\,E_- \,(X \cdot q)\,, \quad 
c_{[26]}^{(p)}= (\hat w \cdot q)\,E_-\,,\quad 
c_{[23]}^{(p)}= \frac{i}{\sqrt{3}}\,E_-\,(X \cdot q)\,,
\nonumber\\
&&c_{[27]}^{(p)}=c_{[18]}^{(p)}= -\sqrt{\frac{3}{2}}\,\Big(\frac{1}{3}\,E_+ \,E_- 
+(X \cdot q)^2 \Big)\,,
\end{eqnarray}
and
\begin{eqnarray}
&&c_{[33]}^{(p)}= \frac{1}{3}\,M^2_-\,E_+\,,\quad 
c_{[44]}^{(p)}=\frac{1}{3}\,E_+^2\,E_- \;,
\nonumber\\
&&c_{[55]}^{(p)}= E_+\,(\hat w\cdot q)^2\,,\quad 
c_{[77]}^{(p)}=\frac{1}{2}\,E_+\,\Big(\frac{1}{3}\,E_+\,E_-
-\big( X\cdot q\big)^2 \Big)\;,
\nonumber\\
&&c_{[66]}^{(p)}= E_-\,(\hat w\cdot q)^2\;,\quad 
c_{[88]}^{(p)}=\frac{1}{2}\,E_-\,\Big(\frac{1}{3}\,E_+\,E_-
-\big( X\cdot q\big)^2 \Big)\,,
\nonumber\\
&&c_{[35]}^{(p)}=c_{[46]}^{(p)}= -\frac{1}{\sqrt{3}}\,(\hat w\cdot q)\,E_+\,E_-
\,,\quad
c_{[57]}^{(p)}= -i\,\sqrt{\frac{2}{3}}\,(X \cdot q) \,(\hat w\cdot q)\,E_+\,,
\nonumber\\
&&c_{[37]}^{(p)}=c_{[48]}^{(p)}= i\,\frac{\sqrt{2}}{3}\,(X\cdot q)\,E_+\,E_- \,,\quad 
c_{[68]}^{(p)}= -i\,\sqrt{\frac{2}{3}}\,(X \cdot q) \,(\hat w\cdot q)\,E_- \;,
\nonumber\\
&&c_{[34]}^{(p)}=  -\frac{i}{3}\,(X \cdot q)\,E_+\,E_-\,,\quad
c_{[56]}^{(p)}= -i\,(\hat w \cdot q)^2 \, (X \cdot q) \;,\;\;\;\;
\nonumber\\
&&c_{[78]}^{(p)}= i\,\big( X\cdot q\big)\left( \frac{3}{2}\, \big( X\cdot q\big)^2
+\frac{5}{6}\,E_+\,E_- \right)\,, 
\nonumber\\
&&c_{[36]}^{(p)}= \frac{i}{\sqrt{3}}\,(\hat w \cdot q)\,E_- \,(X \cdot q) \,,\quad
c_{[38]}^{(p)}=\frac{1}{\sqrt{2}}\,E_- \,
\Big(\frac{1}{3}\,E_+\,E_- +(X \cdot q)^2\Big) \;,
\nonumber\\
&&c_{[45]}^{(p)}= \frac{i}{\sqrt{3}}\,(\hat w \cdot q)\,E_+ \,(X \cdot q) \;,\quad 
c_{[47]}^{(p)}=\frac{1}{\sqrt{2}}\,E_+ \,
\Big(\frac{1}{3}\,E_+\,E_- +(X \cdot q)^2\Big) \;,\quad 
\nonumber\\
&&c_{[58]}^{(p)}=c_{[67]}^{(p)}=-\sqrt{\frac{3}{2}}\,(\hat w \cdot q) \,\Big(
\frac{1}{3}\,E_+\,E_- +(X \cdot q)^2\Big)\,,
\label{}
\end{eqnarray}
where
\begin{eqnarray}
&& E_\pm \equiv m_N\pm (\sqrt{w_0^2-\vec w\,^2}-q\cdot \hat w ) \;, \quad 
E_+\,E_- =q^2-(q\cdot \hat w)^2 \;.
\end{eqnarray}

\end{document}